\RequirePackage{fix-cm}
\documentclass[smallextended]{svjour3}       
\smartqed  
\usepackage{graphicx}
\usepackage{xcolor}
\usepackage{blindtext}
\usepackage[utf8]{inputenc}
\usepackage{natbib}
\usepackage{graphicx}
\usepackage{adjustbox}
\usepackage{rotating}
\usepackage{pdflscape}
\usepackage{url}
\usepackage{amsmath}
\usepackage{multirow}
\usepackage{lipsum}       
\usepackage{changepage}
\usepackage{enumitem}
\usepackage{color}
\usepackage{inputenc}
\usepackage{pgfplots}
\usepackage{amsfonts}

\usepackage{multirow}

\usepackage{amssymb }
\usepackage{comment}

\usepackage[T1]{fontenc}
\usepackage{amsmath}
\usepackage{amsfonts}
\usepackage{amssymb}
\usepackage{footnote}
\usepackage{caption}
\usepackage{subcaption}
\usepackage{adjustbox}
\usepackage{mathtools}
\usepackage{booktabs}

\makeatletter
\let\cl@chapter\undefined
\makeatletter
\usepackage{cleveref}

\begin{document}

\title{Explaining Recommendation System Using Counterfactual Textual Explanations}


\author{Niloofar Ranjbar        \and
        Saeedeh Momtazi* 
        \and
        MohammadMehdi Homayoonpour
}


\institute{Niloofar Ranjbar \at
Computer Engineering Department\\
Amirkabir University of Technology (Tehran Polytechnic) \\
\email{nranjbar@aut.ac.ir}
           \and
             Saeedeh Momtazi \at
Computer Engineering Department\\
Amirkabir University of Technology (Tehran Polytechnic) \\	
             \email{ momtazi@aut.ac.ir}   
             \and
              MohammadMehdi Homayounpour \at
             Computer Engineering Department\\
             Amirkabir University of Technology (Tehran Polytechnic) \\	
             \email{ homayoun@aut.ac.ir}   
}


\maketitle

\begin{abstract}
Currently, there is a significant amount of research being conducted in the field of artificial intelligence to improve the explainability and interpretability of deep learning models. It is found that if end-users understand the reason for the production of some output, it is easier to trust the system. Recommender systems are one example of systems that great efforts have been conducted to make their output more explainable. One method for producing a more explainable output is using counterfactual reasoning, which involves altering minimal features to generate a counterfactual item that results in changing the output of the system. This process allows the identification of input features that have a significant impact on the desired output, leading to effective explanations. In this paper, we present a method for generating counterfactual explanations for both tabular and textual features. We evaluated the performance of our proposed method on three real-world datasets and demonstrated a +5\% improvement on finding effective features (based on model-based measures) compared to the baseline method.
\end{abstract}

\section{Introduction}
Deep neural models are commonly used in a variety of tasks, such as healthcare, decision support systems, and credit risk assessments. However, there is a growing need to ensure the trustworthiness and reliability of these models to make fair and robust decisions.

Recommendation systems aim to predict a score that a user would give to an item, and then suggest the top-ranked items to the user. It is difficult for users to understand why the system is suggesting a particular item, which can make it difficult for them to trust the system's recommendations. Additionally, understanding the reasoning behind a recommendation can help developers debug the system. As a result, this motivated researchers to focus on the explainability of recommendation system outputs.

In this paper, we propose a feature-based method for predicting item scores using a matrix of features. For users, this matrix is based on the items they have interacted with. We then use a deep neural model to predict the score that a new user would give to an item based on these features.

Some methods extract different aspects of an item from user reviews and use these aspects as features. However, as many features may not be immediately mentioned in user reviews, we also utilize metadata from the items' descriptions to extract additional features. This allows for a combination of continuous, categorical, and textual features to be used for training the model.

Inspired by counterfactual explanation generation methods, we introduce methods that use cost functions to identify and highlight the features that have the greatest impact on the predicted item score. For example, if a particular mobile phone is predicted to have a high score, these methods can identify the features that contributed to this high score, such as the model, battery, RAM, and camera. In one of these methods, we employ the Gumbel Softmax trick , allowing us to consider all types of important features (continuous, categorical, and textual) simultaneously. To the best of our knowledge, this is the first time that all types of features have been simultaneously included in counterfactual explanations.

It is worth noting that we have chosen to use counterfactual explanation generation methods, as they have been shown to be more understandable to humans and more closely align with human thought processes, according to \cite{yang2020generating}.

Our paper makes several contributions to the field of recommender systems:
\begin{itemize}
\item We introduce a novel feature-based method for predicting item scores, which allows for a combination of continuous, categorical, and textual features to be used for training. 
\item We modify the counterfactual explanation generation method proposed by \cite{tan2021counterfactual} to enable it to generate explanations when raw text is used as a feature in a recommender system.
 \item We propose a new counterfactual explanation generation method that utilizes a genetic algorithm to generate explanations when raw text is used as a feature.
 \item We introduce a counterfactual explanation generation method based on the Gumbel-softmax method, which can generate explanations when all types of features (continuous, categorical, and textual) are used in a recommender system.
\end{itemize}
These contributions can have practical applications in various domains.

The paper is structured as follows: in Section \ref{relatedWorks}, we review related works in counterfactual explanations and explainable recommendations.
In Section \ref{expgenmethods}, we introduce our three explanation generation methods, which include CountER, Genetic algorithm, and Gumbel-Softmax based method.
In Section \ref{experiments}, we describe the experiments we conducted to evaluate the different methods. We then present the results of our experiments on the Amazon and Yelp datasets.
Finally, in Section \ref{conclusion}, we conclude with a discussion of our findings and suggestions for future work.

\section{Related Works}\label{relatedWorks}
As we introduce a counterfactual explanation method and test it for a recommendation system, we divide the related works into two categories: counterfactual explanation methods applied to text, and explainable recommendation systems. Our method represents a novel combination of counterfactual explanations and recommendation systems, and we expect it to be of interest to researchers and practitioners in both fields.

\subsection{Related works in Counterfactual explanations}
In simple terms, a counterfactual explanation for a prediction identifies the smallest alteration that can be made to the input features to produce a desired or predefined outcome instead of the predicted one\citep{molnar2022}.
There have been many efforts to apply counterfactual explanations to textual data. In such cases, the explanations should appear natural to humans. Simply removing words from the text to generate counterfactual explanations is not effective. \cite{yang2020generating}  addressed this issue by ensuring that replaced words are grammatically correct. They demonstrate their approach on a sentiment analysis task, introducing two lists of words: one containing words that are suitable for replacement based on grammar, and another containing words with opposite senses to those in the sentiment dictionary. They then identified the intersection of these two lists and replaced words in the main text with words from this intersection until the predicted class was changed. This approach helps to generate counterfactual explanations that are more understandable to humans.

Many works have addressed the issue of generating natural-sounding counterfactual explanations in text using language representation models such as BERT \citep{devlin2018bert}. \cite{fern2021text} proposed one example of such an approach. They first generated a candidate set of words to replace each word in the text. They then used BERT as a language model to determine the probability of each candidate token for a given position. In the second step, they found the best combination of changes using shapley values \citep{kalai1987weighted} and generated the explanations using beam search. This approach allows the generation of more coherent and understandable counterfactual explanations in text.

The proposed models by \cite{madaan2021generate} and \cite{wu2021polyjuice} both generated counterfactual explanations in a conditional manner using GPT2. \cite{madaan2021generate} defines named-entity tags, semantic role labels, or sentiments as conditions for the words, while \cite{wu2021polyjuice} controls the types and locations of perturbations in the text. Both approaches allow more targeted and controlled generation of counterfactual explanations in text.

Using pre-trained language models to generate alternative texts as adversarial examples for text has become a popular approach in recent years. As the goal of generating adversarial examples is similar to that of generating counterfactual explanations (i.e., minimally changing the input text to change the prediction class), works in this direction can be considered related to our approach. \cite{guo2021gradient} attempted to generate the most probable sentence using BERT as a language model, while \cite{garg2020bae} and  \cite{li2020bert} used BERT to suggest word replacements. These works demonstrate the utility of language models in generating alternative text that can fool prediction models.

There are many tasks, such as recommendations and healthcare, that involve hybrid data comprising text, continuous features, and categorical features. To generate counterfactual explanations for this type of data, a method that can handle all data types simultaneously is needed. None of the previously mentioned methods address this issue. In this paper, we propose a method to address this issue and generate counterfactual explanations for hybrid data.

\subsection{Related works in Explainable Recommendations}
There have been numerous efforts to make recommender systems more explainable. One approach involves extracting various aspects of items from user reviews and using them as input features for black-box models. After training the black-box model, the goal is to identify the minimal set of features that are most important in ranking items for a particular user. \cite{zhang2014explicit} employed matrix factorization to achieve this, while \cite{chen2016learning} and \cite{wang2018explainable} used tensor factorization instead. Other works have utilized counterfactual reasoning to generate explanations. For example, \cite{zhou2021intrinsic} employed three strategies to make white-box, gray-box, and black-box models explainable by using attention weights, adversarial and counterfactual perturbations, and extracting aspects from user reviews as features. \cite{ghazimatin2020prince} attempted to find the minimal set of user actions, such as ratings, that would cause a recommended item to change when removed. \cite{tan2021counterfactual} introduced an aspect-based recommender system and attempted to make it explainable through the use of counterfactual reasoning by trying to solve a joint optimization problem that minimally changes item aspects such that the new item is no longer recommended. \cite{pan2021explainable} attempted to map uninterpretable features to interpretable ones by minimizing both prediction and interpretation loss.

On the other hand, \cite{wang2018tem} argued that tree-based models are interpretable and neural-based models have acceptable results in recommender systems. Therefore, proposed combining these two types of models to create an explainable recommender system.

Other works have utilized knowledge graphs to make recommender systems more explainable. For example, \cite{wang2020learning} attempted to represent knowledge-graph paths with the semantic information of entities and their relations in order to make recommendations generated by the knowledge graph more explainable. \cite{syed2022context} used first-order logic to generate triples from users' complex queries and then tried to find entities that satisfy these logical queries using a knowledge graph. These entities were sorted based on the information they captured from the context, and explanations were generated using the triples. \cite{shimizu2022explainable} introduced a knowledge graph attention network that used side information of items to make recommendations and generate explanations. \cite{geng2022path} trained a language model on the paths of a knowledge graph consisting of entities and edges, which were based on user actions and item features as well as the relationships between them. Explanations were generated using the resulting graph.

While many of the explanations generated through these methods can help users understand why an item is recommended to them, they do not necessarily assist sellers and managers in better satisfying their users. Counterfactual explanations, on the other hand, can provide sellers and managers with information about which features of items need to be changed in order to more effectively recommend them to users. In this paper, we utilize counterfactual reasoning to generate accurate, trustworthy, and comprehensive explanations.

\section{Explanation generation methods} \label{expgenmethods}
In this section we introduce three methods to generate explanations.
\subsection{CountER} 
\label{countER}
Inspired by the work of \cite{tan2021counterfactual}, we sought to identify a slight change vector to add to the weights of textual feature vectors when averaging them. In this section, we first discuss the base Counterfactual method, and then propose our own method.
\subsubsection{The base CountER model}
Suppose we have a set of \textit{m} users, $\mathcal{U}=\{u_1,u_2,...,u_m\}$, and a set of \textit{n} items, $\mathcal{V}=\{v_1,v_2,...,v_n\}$. For each type of item (e.g., cell phones), we extract a list of \textit{r} aspects, $\mathcal{A}=\{a_1,a_2,...,a_r\}$. We then construct the user-aspect preference matrix, $\mathcal{X} \in\mathbb{R}^{m\times r}$, and the item-aspect quality matrix, $\mathcal{Y} \in\mathbb{R}^{n\times r}$, using scores calculated as follows:
\begin{equation}
	\begin{aligned}
		\mathcal{X}_{i,k}=
		\begin{cases}
			0 ,\text{ if user $u_i$ did not mentioned aspect $a_k$}\\
			1+\left(N-1\right)\left(\frac{2}{1+exp(-t_{i,k})}-1\right) , \text{ otherwise}
		\end{cases} \\    
		\mathcal{Y}_{j,k}=
		\begin{cases}
			0 , \text{ if item $v_j$ is not reviewed on aspect $a_k$}\\
			1+\left(\frac{N-1}{1+exp(-t_{j,k}.s_{j,k})}\right) ,\text{ otherwise}\\
		\end{cases}
	\end{aligned}
\end{equation}
\label{eq2}
In this context, \textit{N} represents the rating scale, which is set to 5 in this case. $t_{i,k}$ denotes the frequency with which user $u_i$ mentions aspect $a_k$, $t_{j,k}$ denotes the frequency with which aspect $a_k$ is mentioned in item $v_j$ reviews, and $s_{j,k}$ represents the average sentiment of these mentions.\\
After training the black-box model, researchers identify the most effective aspects by solving the following optimization problem:
\begin{equation}
	\begin{aligned}
		\text{minimize C}(\Delta)=\lVert \Delta \rVert_2 ^2 + \gamma \lVert \Delta \rVert_0  \\ 
		\text{s.t. S}(\Delta)=s_{i,j\Delta} \leq s_{i,j K+1}
	\end{aligned}
\end{equation}
$\Delta=\{\delta_0, \delta_1, ... , \delta_r\}$ is a vector with zero or negative values. In this optimization problem, $\Delta$ is learned for item $v_j$ such that by applying it on the $\mathcal{Y}_j$ vector ($\mathcal{Y}_{j+\Delta}$), the item will no longer be in the top \textit{K} list of items recommended to user $u_i$. The term $\lVert \Delta \rVert_2 ^2 $ controls the amount of change in the $\Delta$ values, while $\lVert \Delta \rVert_0$ controls the number of values that change in $\Delta$ vector. $s_{i,jK+1}$ is the ranking score of the marginal item (the last item in the top K list of recommended items), and $s_{i,j\Delta}$ is the ranking score of item $v_j$ after applying $\Delta$ to its aspect vector.

By optimizing this equation, $\Delta$ is learned such that the item's aspect vector is minimally changed until it is removed from the top \textit{K} recommended items for user $u_i$.

Finally, the aspects that have negative values in $\Delta$ are important, and removing them from the list of aspects causes a change in the system's recommendation decisions.

Since the terms $\lVert \Delta \rVert_0$ and $s_{i,j\Delta} \leq s_{i,j K+1}$ are not differentiable, they choose $\lVert \Delta \rVert_1$ instead of $\lVert \Delta \rVert_0$ and relax $s_{i,j\Delta} \leq s_{i,j K+1}$ as a hinge loss. Therefore, the final optimization equation is as follows:
\begin{equation}
	\label{eq3}
	\begin{aligned}
		\underset{\Delta}{minimize}\lVert \Delta \rVert_2 ^2 + \gamma \lVert \Delta \rVert_1 + \lambda max(0,\alpha +s_{i,j\Delta} - s_{i,j K+1} )
	\end{aligned}
\end{equation}
where $a=0.2$, $\lambda=100$ and $\gamma=1$ based on the proposed model by \cite{tan2021counterfactual}.
\subsubsection{CountER for word vectors}
To determine which words in the text are more important and critical to the model, we attempt to learn a vector $\Delta=\{\delta_0,\delta_1, ..., \delta_z\}$ as in the base CountER model and add it to the basic weight vector $\Theta=\{1,1,...,1\}$, for a list of word features $W=\{w_0,w_1, ..., w_z\}$ of item $v_j$. The basic weight vector assigns a weight of 1 to every word $w_t$ in the list of item $v_j$ features. At the end, words that have a weight less than a threshold $t$ are considered important features, so that removing them causes the item to no longer be in the top \textit{K} list.

Since the existence of words in the features is a binary decision, we remove the term $\lVert \Delta \rVert_2 ^2$ from Equation \ref{eq3}, as the amount of change is not important in this case.
\subsection{Genetic algorithm} 
As previously stated, the inclusion of words in the features is a binary decision, and thus, we introduce another algorithm based on genetic algorithms. The Genetic Algorithm in its conventional form employs a collection of potential solutions that act as representations of a resolution to the optimization problem that requires solving \citep{kramer2017genetic}. For a list of word features $W=\{w_0,w_1, ..., w_z\}$ for item $v_j$, we define chromosomes as binary vectors of size z. The steps of the algorithm are as follows:
\begin{itemize}
	\item{\textbf{Making random population: } We considered population sizes of $\in\{100,200,400\} $. The first population was generated randomly, with a probability of 0.9 for the number of ones and 0.1 for the number of zeros in each chromosome, as we aim to minimize the number of removed words.
	}
	\item{\textbf{Selection: } In order to select chromosomes for the next generation, we first calculate the fitness score for each chromosome. The probability of a chromosome being chosen is based on its fitness score, where chromosomes with higher fitness scores are more likely to be selected. For a chromosome \textit{c} with size \textit{z}, we calculate the fitness as follows:
		\begin{equation}
			\begin{aligned}
				fitness_{c}= \frac{1}{\lambda\times(\alpha + s_{i,jc} - s_{i,jK+1})} + countScore_c \\
				countScore_c=
				\begin{cases}
					0.5 \times (1-\frac{\sum_{r:c_r\neq 0}  1}{z}) & s_{i,jc} > s_{i,jK+1}\\
					\beta \times (\frac{\sum_{r:c_r\neq 0}  1}{z}) &\text{otherwise}
				\end{cases} \\    
			\end{aligned}
		\end{equation}
		
		where $s_{i,jc}$ is the ranking score of item $v_j$ after applying the values of chromosome c as weights to its main vector, and $s_{i,jK+1}$ is the ranking score of the marginal item.
		As can be seen, the fitness function has two parts. The first part increases when the score of the item decreases. $\alpha= 1$ is added to the difference of scores to prevent negative values. The second part of the fitness function $(countScore_c)$ is a penalty term that aims to minimize the number of removed features. When the ranking score of the item is greater than the marginal ranking score, some features need to be removed to decrease the ranking score, so in this situation, removing more features is better, but we consider a low coefficient (0.5) for that. On the other hand, when the ranking score of the item is less than the marginal ranking score, it means that the item is no longer in the top-K items and the goal has been met. Therefore the number of removed features should be minimized as much as possible for this model. $\lambda$ and $\beta$ are hyperparameters that need to be tuned.
	}
	\item{\textbf{Cross over: } After the selection phase, it is time to perform crossover on pairs of chromosomes. Crossover is done with a rate of 99\%. 
	}
	
	\item{\textbf{Mutation: } Mutation is performed with a rate of 10\% on at most 50\% of the population. For each chromosome, at most 10\% of its genes are changed.}
	
\end{itemize}
Steps 2 to 4 are repeated for 10 iterations if the best fitness score exceeds a threshold (1 is chosen based on experiments), otherwise, the algorithm continues for up to 50 iterations to allow for the possibility of finding a better solution.
\subsection{Gumbel-Softmax based method} 
In counterfactual explanations, we need to make minimal changes to the features such that the recommended item is no longer in the top \textit{K} list. For continuous features, this is straightforward, but for textual features, changing them is equivalent to removing them or replacing them with other words. Removing words from text can produce meaningless sentences. Therefore replacing them with other words is a better approach. This is also true for categorical features, which must have predefined values. Since they cannot be removed from features, changing them is equivalent to choosing a new value from a predefined set.\\For optimization problems where the parameters are discrete, such as this one, a new gradient estimator called Gumbel-Softmax has been introduced by \cite{jang2016categorical}. It is based on Gumbel-Softmax distribution and is a powerful gradient estimator that swaps out the non-differentiable sample from a categorical distribution with a differentiable sample from a Gumbel-Softmax distribution. This distribution's crucial characteristic is its ability to smoothly anneal into a categorical distribution. \cite{guo2021gradient} used this trick to generate new meaningful text, we use it to find the best alternative words. In the following, we describe the method we use for textual features. As the textual features are similar to categorical features, we can use this method for categorical features as well.

Suppose we have a list of words (as textual features) $w= \{w_0,w_1,...,w_z\}$ for item $v_j$. Changing these words is equivalent to replacing them with other probable words such that the meaning of the whole sentence does not change much. We find the five most probable alternative words for each of them using BERT as a language model. We mask a word and make BERT predict alternative words.

Consider a distribution $P_\Theta$ parameterized by a matrix $\Theta \in\mathbb{R}^{z\times (5\times z)}$. For each word $w_r$, we have a vector of token probabilities in the $\Theta$ matrix named $\pi_r$, where $\pi_{r}= Softmax(\Theta_{r})$. At first, we choose the probabilities such that for each word, the word itself is chosen as the alternative word. After that, the parameter matrix $\Theta$ is optimized such that the score of the item decreases with minimum replacements in the words. As the Softmax function is not differentiable, the Gumbel-Softmax approximation is used. Samples from the Gumbel-Softmax distribution $\tilde{P}_\Theta$ are drawn as follows:
\begin{equation}
	(\tilde{\pi}_{r})_{k} := \frac{exp((\Theta_{r,k}+g_{r,k})/T)}{\sum_{v=1}^{V}exp((\Theta_{r,v}+g_{r,v})/T) }
\end{equation}
where $(\tilde{\pi}_{r})_{k}$ is the $k^{th}$ value of the vector $\tilde{\pi}_{r}$, $g_{r,k}\sim Gumbel(0,1)$, $V=5\times z$ is the size of the alternative words list, and $T>0$ is a temperature parameter that controls the smoothness of the Gumbel-softmax distribution. When $T\rightarrow 0$ this distribution converges to a categorical distribution.

To find the contextual vector for each alternative word, we replace the main word in the sentence with the alternative word and then extract the contextual word vector as will be discussed in Section \ref{amazondat}. Finally, we have a matrix $C \in \mathbb{R}^{V \times 768}$ such that for each alternative word, we have a vector of size 768. By multiplying the $\pi \in\mathbb{R}^{z\times V}$ matrix by the $C\in \mathbb{R}^{V \times 768}$ matrix, we have a matrix with dimensions $z\times 768$. Therefore for each word $w_r$ in $w=\{w_0,w_1,...,w_z\}$ we have a vector of size 768. The optimization equation defined in equation \ref{eq3} changes as follows:
\begin{equation}
	\label{eq6}
	\begin{aligned}
		\underset{\Theta}{minimize}\text{ }\lambda max(0,\alpha +s_{i,j\Theta} - s_{i,j K+1} )
	\end{aligned}
\end{equation}
where $s_{i,j,\Theta}$ is the predicted score of item $v_i$ for user $u_j$ after using $\Theta$ to calculate $\pi$ values and then applying $\pi$ on the word vectors, as previously mentioned.
In order to replace more probable words with each word, we define a matrix $L \in \mathbb{R}^{z \times V}$. For each $l_{i,k} \in L$, if the word $a_k$ is in the alternative list of the word $w_i$, $l_{i,k}$ equals the BERT output layer logit for the word $a_k$, otherwise it equals -1. Note that higher values of logits indicate higher ranks in the top 5 alternative words list.
Furthermore, we add the $l_1$ norm of the difference between the main $\pi$ calculated with the main words (no words are replaced) and the $\pi$ calculated while optimizing the $\Theta$ matrix, as the penalty term for the number of replaced words.
Therefore, the final optimization equation is as follows:
\begin{equation}
	\label{eq7}
	\begin{aligned}
		\underset{\Theta}{minimize}\text{ }  \lambda max(0,\alpha +s_{i,j\Theta} - s_{i,j K+1} )+ \beta \frac{1.0}{\tilde{\pi}\cdot L}+ \gamma \lVert \tilde{\pi}_{main}-\tilde{\pi} \rVert_1 
	\end{aligned}
\end{equation}
where $\alpha=0.2$, $\lambda=100$ (as stated in Section \ref{countER}), $\beta$ and $\gamma$ are hyperparameters, and $\tilde{\pi} \cdot L$ is the dot product of $\tilde{\pi}$ and $L$. It is worth noting that as the model chooses more probable words, this dot product increases.

For categorical features, we use the same formulation, except that the alternative values for each categorical feature are not words anymore. Additionally, there is no need to calculate the second term in equation \ref{eq7}. Finally, for a combination of continuous, categorical and textual features the optimization problem is as follows:
\begin{equation}
	\label{eqfinal1}
	\begin{aligned}
		\underset{\Delta, \Theta1 ,\Theta2 ,\Theta3 }{minimize}\text{ } l_{score}+l_{textual}+l_{cat}+l_{continuous}\\
		l_{score}= \lambda max(0,\alpha +s_{i,j_{\Delta, \Theta1 ,\Theta2 ,\Theta3} } - s_{i,jK+1} )\\
		l_{textual}=\beta \frac{1.0}{\tilde{\pi_1}\cdot L}+ \gamma \lVert \tilde{\pi}_{main_1}-\tilde{\pi}_{1} \rVert_1 \\
		l_{cat}=\gamma \lVert \tilde{\pi}_{main_2}-\tilde{\pi}_{2} \rVert_1\\
		l_{continuous}=\lVert \Delta \rVert_2
   \end{aligned}
\end{equation}
Here, $l_{score}$ is the same as the loss explained in equation \ref{eq6}, with one main difference: all types of features are used to calculate the new score. Therefore, all values of ${\Delta, \Theta1 ,\Theta2 ,\Theta3}$ are used to calculate this score. $l_{textual}$ is the same as the one defined in equation \ref{eq7}, in which $\Theta1$ is used to calculate $\tilde{\pi}_{1}$. $l_{cat}$ is calculated for all categorical features, in which $\Theta2$ is used to calculate $\tilde{\pi}_{2}$. Finally, $l_{continuous}$ is used to ensure that the $\Delta$ used for continuous features changes minimally.
\section{Experiments}\label{experiments}
In this section, we first specify datasets and settings needed for the experiments. Next, we evaluate the introduced methods both quantitatively and qualitatively. Finally, we discuss the results.  Note that all experiments were conducted on NVIDIA T4 Tensor Core GPUs using Colab.
\subsection{Datasets}
We test our methods on the Amazon and Yelp datasets. The Amazon dataset contains 29 sub-datasets, each of them for a specific product category. We use two sub-datasets of different scales, "Cell Phones and Accessories" and "CDs and Vinyl". Each dataset contains user reviews of items, item descriptions, and some additional features as textual data. The Yelp dataset, on the other hand, contains information on various businesses such as their geographic location (latitude and longitude), opening hours, and other special features (e.g., parking, food types for restaurants, etc.). It also includes user reviews and tips on the businesses. We consider longitude and latitude as continuous features, tips as textual features, and some other features as categorical features.

Table \ref{tab:datasets} shows the statistics of the datasets. For all datasets, we drop users and items with less than 5 and 10 reviews, respectively. To prepare the train, validation and test sets, we only use users with more than 15 items interacted with. For each user, these items are considered as positive samples. We hold out the last 10 items for validation and test sets (5 each) and use the remaining items for the train set. We also sample negative instances randomly from the non-interacted items with a ratio of 1:5, meaning for each positive instance, we sample five negative instances. 
\begin{table}[]
	\caption{Statistics of the datasets}
	\label{tab:datasets}
	\begin{center}
		\begin{tabular}{ccccccc}
			\toprule
			Dataset & Users & Items & Reviews & Tips & Train & Test\\
			\midrule
			Cell Phones and Accessories & 6794 & 1945 & 36762 & N/A & 2871 & 93\\
			CDs and Vinyl & 11467 & 11677 & 224090 & N/A & 686172 & 3631\\
			Yelp & 243531 & 48089 & 1426442 & 626069 & 501606 & 3529\\
			\bottomrule
		\end{tabular}
	\end{center}
\end{table}
\subsection{Data preparation}
\subsubsection{Preparing Amazon dataset}
\label{amazondat}
\begin{figure}[h]
	\centering
	\includegraphics[width=\linewidth]{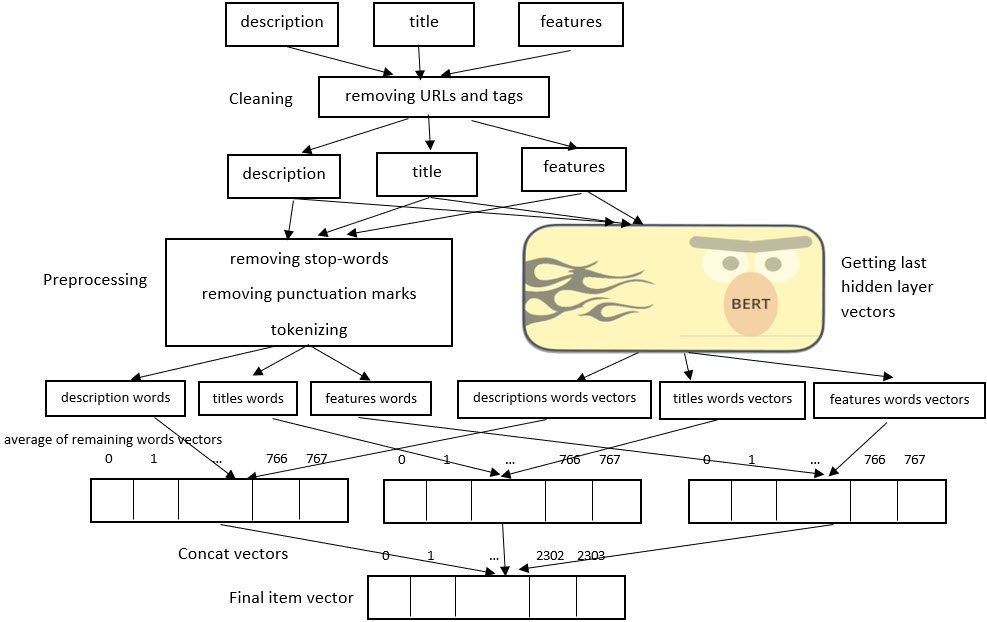}
	\caption{Preparing amazon data for black-box model}
	\label{img:amazondata}
\end{figure}

As can be seen in Figure \ref{img:amazondata} for amazon datasets we extract textual features from descriptions, titles and features of items by using the BERT. The steps to extract textual features are as follows:
\itemize{
	\item{\textbf{Cleaning texts:} We remove URLs and some xml tags like ``</b>" and the text between them. We remove words which are the combination of digits and characters, because there are many meaningless words like them in the text.}
	\item{\textbf{Extracting contextual vectors using BERT:} We use ``bert\_base\_uncased" which produces a vector of size 768 for each word in each hidden layer. To extract contextualized vectors for the words in the sentences, we use the last hidden layer of the BERT.}
	\item{\textbf{Preprocessing:} To extract valuable words from the sentences we remove stop-words and punctuation marks from the texts.
	}
	\item{\textbf{Generating final item vector:} We calculate the average of the remaining word vectors and then concatenate descriptions, titles and features vectors to get a vector of size 2304 as the final item vector. Note that for the words which are not in the BERT vocabulary and BERT tokenizes them into several parts, we calculate the average of all parts vectors as the word vector.
	}
}
\subsubsection{Preparing Yelp dataset}
\label{yelp-prep}
The proposed method was tested on the Yelp dataset, which contains a combination of textual, continuous, and categorical features. Textual features were extracted from user tips on businesses, and vectors were generated as previously described for the Amazon dataset. Continuous features included the latitude and longitude of each business, while categorical features included information such as the time periods during which the business is open, whether it has parking, and the types of food it serves. A binary vector was created to represent the time periods, with a value of 1 indicating that the business is open during that hour and a value of 0 otherwise. The 87 categorical features were encoded with three values: -1 for businesses that do not have the feature, 0 for businesses for which it is not specified whether or not they have the feature, and 1 for businesses that have the feature. All continuous and categorical features were then scaled to have zero mean and unit variance. The final vector for each business was obtained by concatenating the textual features vector with the scaled continuous and categorical features, resulting in a vector of size 168+87+768=1023.
\subsection{setup of experiments}
\subsubsection{Black-box model}
The black-box model is a feed forward network with 2 hidden layers containing 512 and 256 neurons. The input to the model is a concatenation of two vectors. The first vector is for the item and the second one is for the user. Any kind of textual features, as well as continuous and categorical features can be used in this model based on their availability. Note that we generate the user vector by averaging the items vectors he/she interacted with.

We apply the ReLU activation function after each layer except the last one, in which we use a Sigmoid activation function. The output layer maps the ranking score, $s_{i,j}$, to a value within the range of (0,1), allowing us to recommend top-K items for a user based on the predicted ranking scores.

A cross-entropy loss is used for training the model, and we employ a stochastic gradient descent (SGD) optimizer with a learning rate of 0.01.
\subsubsection{Hyper-parameters}
A stochastic gradient descent (SGD) optimizer with a learning rate $(lr)$ is employed to optimize all methods, with the exception of the genetic method. A value of $lr=0.01$ is chosen for the countER methods, and it is tuned for the Gumbel-softmax-based method.

As previously discussed in Section \ref{countER}, for the main countER method, we set the hyper-parameters in Equation \ref{eq3} to the same values as those used by \cite{tan2021counterfactual}. However, for the countER for word vectors, we tested different values for $\gamma$ in ${0.2,0.3,0.4,...,1.0}$ and for the threshold $t$ in ${0.01,0.1,0.2,0.3,0.4,0.5}$ on the validation set and found that the best values were $\gamma=0.7$ and $t=0.3$.

For the genetic method, we tested different values for $\lambda$ in ${2,5,10}$ and for $\beta$ in ${2,10,50}$ on the validation set and found that the best values were $\lambda=10$ and $\beta=10$.

The Gumbel-softmax-based method has temperature $T$, learning rate $lr$, $\beta$, and $\gamma$ as hyper-parameters. We test different values for each dataset separately. We tested different values for $T$ in $\{0.5,1.0,1.2,1.4,1.5,2.0,2.2,2.4\}$ (best= $2.0$ and $2.2$), for $lr$ in $\{0.1,0.2,...,0.7\}$, (best= $0.5$ and $0.7$) for $\beta$ in $\{1000,2000,10000\}$ (best= $1000$) and for $\gamma$ in $\{1,2,5\}$ (best= $1$).
\subsubsection{Evaluation Metrics}
We use NDCG (Normalized Discounted Cumulative Gain) as a metric for measuring the ranking quality of the black-box model. 

Inspired by \cite{tan2021counterfactual}, we evaluate the explanation generation methods from two perspectives: user-oriented and model-oriented evaluations.
\begin{itemize}
	\item{\textbf{user-oriented evaluation:} In user-oriented evaluation, we examine the extent to which the features of an item extracted by the explanation generation method are mentioned in the user's reviews of that item.
		
		Therefore, for an item $v_j$ and user $u_i$, if the explanation generation method outputs $E={e_1,e_2,...,e_N}$ as the important features, and the user $u_i$ mentions the words $G={g_1,g_2,..., g_M}$ in their review of item $v_j$, the precision, recall and F1 metrics for the user-oriented evaluation are calculated as follows:
		\begin{equation}
			\label{eq8}
			\begin{aligned}
				Precision=\frac{\sum_{k=1}^{r}p_{i,j}^{k}}{N}, 
				Recall=\frac{\sum_{k=1}^{r}p_{i,j}^{k}}{M}\\
				F1=2\cdot\frac{Precision\cdot Recall}{Precision+Recall}\\
				p_{i,j}^{k}=
				\begin{cases}
					1 & e_{i,j}^k\in G_{i,j}\\
					0 &\text{ otherwise}
				\end{cases} 
			\end{aligned}
		\end{equation}
		where N is the number of features chosen by the explanation generation method and M is the number of words used by the user in their review of the item. Finally, we average the scores of all pairs to obtain the final precision, recall and F1. It should be noted that before using the user's reviews as ground-truth features, we preprocess them by removing punctuation marks and stop-words. The review words are then tokenized and used as the ground-truth features.}
	\item{\textbf{model-oriented evaluation: } In user-oriented evaluation, we determine whether the generated explanation features are consistent with the user's preferences. To check whether the generated explanation features truly capture the model's behavior, we use two additional metrics, Probability of Necessity (PN) and Probability of Sufficiency (PS).
		
		The PN metric answers the question "Are the generated features necessary for the model to predict the rank correctly?" To answer this question, we remove the specified features from the list of features, and check whether the item is still recommended to the user.
		
		The PS metric answers the question "Are the generated features sufficient for the model to predict the rank correctly?" To answer this question, we use only the specified features and remove other features from the list of features and check whether the item is still recommended to the user.
		
		We calculate the harmonic mean of PN and PS as a third metric named FNS, similar to the F1 measure.}
\end{itemize}
Another important metric in evaluating explanation generation methods is their stability. This metric measures how consistent the explanation features generated by the model are across different runs. We define the stability of the model for an item $v_j$ and user $u_i$ as follows:
\begin{equation}
	\label{eq9}
	Stability_{i,j}=\frac{1}{N(N-1)} \sum_{k=1}^{N}{\sum_{l=1 , k \neq l}^{N} {\frac{\lvert r_k \cap r_l\rvert}{\lvert r_k \cup r_l\rvert}}}
\end{equation}

where $r_k$ is the set of explanation features generated by the method in the $k^{th}$ run and N is the number of runs, which is set to 10 for all datasets. The final stability of the model is determined by averaging the scores of all pairs. We calculate the stability for all datasets by using 10 examples from the test set, which are chosen randomly.
\subsubsection{Comparable Baseline}
In our experiments, we used the baseline method proposed by \cite{tan2021counterfactual} as our sole competitor. This method was compared to other state-of-the-art methods in their paper, and it was demonstrated to significantly outperform them. Therefore, we chose to compare our method with the baseline method, which itself outperformed other methods.
\\While we acknowledge that there may be other methods that could potentially be used as competitors, we decided to focus on the baseline method due to its superior performance and the fact that it has already been compared to other competitors in the field. Additionally, given the changes in the dataset, it would not be appropriate to directly compare our results to those of other papers without re-running their codes. Therefore, we chose to re-run the code from \cite{tan2021counterfactual}, which achieved the best results in the field and has already been compared to other competitors. 

\subsection{Amazon dataset Results}
\label{amzresult}
As we use different types of features for the Amazon and Yelp datasets, we present their results separately. The results of our experiments for the Amazon dataset can be seen in Table \ref{tab:amzres} and \ref{tab:amzres2}. It should be noted that the baseline method in these tables refers to the main countER method presented by \cite{tan2021counterfactual}.
\subsubsection{Are generated features based on user preferences?}

As can be seen in Table \ref{tab:amzres}, the value of precision, recall and F1 score of user-based measure decreases for all of the introduced methods compared to the baseline. The reason is as follows: the baseline method uses a fixed set of aspects with 88 and 230 aspects for the cell-phones and CDs datasets respectively. and the model outputs a subset of these aspects as the important features. However, our methods output important words from the item's descriptions, titles and features which are not a specific set of words. Besides, these words may be really important for the user while making a decision but not used in his/her review directly. So this is not fair to compare our methods with the baseline method with this measure. \\
By comparing our methods to each other with this measure, we can find there is no significant difference between them.
\begin{table}[]
	\centering
	\caption{Amazon Dataset Explanation Results, user-based and model-based}
	\label{tab:amzres}
	\resizebox{\textwidth}{!}{%
		\begin{tabular}{c c ccc ccc}
			\hline
			dataset & &\multicolumn{3}{c}{User-based} & \multicolumn{3}{c}{Model-based} \\ 
			& & Pre & Rec & F1 & PN & PS & FNS \\ \hline
			\multirow{4}{*}{cell-phones} & baseline &\textbf{0.273} & \textbf{0.298} & \textbf{0.250} & 0.950 & 0.918 & 0.934  \\ 
			& countERText & 0.1075 & 0.0497 & 0.057 & 0.915 & 0.971 & 0.9426 \\  
			& Genetic & 0.085 & 0.092 & 0.072 & 0.890 & \textbf{0.995} & 0.940  \\ 
			& Gumbel & 0.110 & 0.068 & 0.073 & \textbf{0.970} & 0.990 & \textbf{0.980} \\ \cline{1-8} 
			\multirow{4}{*}{CDs} & baseline & \textbf{0.223} & \textbf{0.329} &\textbf{0.228}  & 0.778 & 0.679 & 0.725 \\ 
			& countERText &0.177 & 0.016 & 0.025 & 0.820 & 0.997 & 0.90 \\ 
			& Genetic &  0.121 & 0.034 & 0.038 & \textbf{0.957} & 0.997 & \textbf{0.97}  \\ 
			& Gumbel & 0.117 & 0.024 & 0.028 & 0.796 & \textbf{0.998} & 0.886 \\ \hline
		\end{tabular}
	}
\end{table}
\begin{table}[]
	\centering
	\caption{Amazon Explanation Results, Other measures}
	\label{tab:amzres2}
		\begin{tabular}{c c c c c c c}
			\hline
			dataset &  & NDCG & Features Avg & Exp Found Rate& Stability & Time(sec) \\ \hline
			\multirow{4}{*}{cell-phones} & baseline & 0.3561 & 2.97 & 78\% & \textbf{0.862} & 1.28  \\
			& countERText & \textbf{0.4047} & 6.05 & 98.9\% & \textbf{0.862} & 1.69 \\  
			& Genetic & \textbf{0.4047}& 15.94 & 98\% & 0.175 & 4 \\ 
			& Gumbel & \textbf{0.4047} & 9.61 & 99\% & 0.51& 7.4 \\ \hline
			\multirow{4}{*}{CDs} & baseline & 0.481 & 5.09 & 72\% & \textbf{0.68} & 1.35\\
			& countERText & \textbf{0.712} & 2.567 & 98\% & 0.65 & 0.43 \\  
			& Genetic & \textbf{0.712} & 7.32  & 95\% & 0.47 & 1.76 \\
			& Gumbel & \textbf{0.712}  & 6.97 & 74\% & 0.54 & 4.8 \\ \hline
		\end{tabular}
\end{table}
\subsubsection{Are generated features specifying the model's behavior?}
The results presented in Table \ref{tab:amzres} demonstrate that our proposed methods, which are based on model-based measures, significantly outperform the baseline.
Upon examination of the data, it becomes apparent that the genetic and Gumbel-softmax algorithms yield the best results for both datasets. Specifically, when analyzing the cell-phone dataset, the Gumbel-softmax algorithm exhibits superior performance in regards to the PN and FNS measures, with the difference in performance for the PS measure being negligible when compared to the Genetic algorithm. On the other hand, when analyzing the cds dataset, the Genetic algorithm demonstrates a marked improvement in PN and FNS measures in comparison to all other methods, while its performance in regards to the PS measure is comparable to that of the other methods.
\subsubsection{The effect of changing features on ranking}
As can be seen in Table \ref{tab:amzres2}, the NDCG value, which measures the ranking quality of the black-box model, increases significantly compared to the baseline for both datasets. This improvement is due to the use of different feature inputs.  However, it should be noted that the three models, countERText, Genetic, and Gumbel, have the same recommendation method, while benefiting from different explanation generation models. Therefore, NDCG shows the same results for all three models. This is because NDCG only determines the accuracy and quality of the recommendation.

Overall, by using these features, it appears that it is possible to recommend items to the user more reliably. The next step is to determine whether the proposed methods are reliable enough to generate accurate explanations based on user preferences.
\subsubsection{Which method finds explanations with lower number of features in average?}
As can be seen in Table \ref{tab:amzres2}, the baseline (countER) and the countERText methods generate explanations with the least number of features for the cell-phones and cds datasets, respectively. The Gumbel-softmax method is the next one in this ranking, and finally, the genetic algorithm finds the most number of features. However, it is worth noting that the genetic algorithm is also able to find more explanations, which may suggest that it is able to find explanations for pairs that other algorithms cannot by using a larger number of features. Another reason may be the difference in the formulation and optimization of the genetic algorithm compared to the other methods.
\subsubsection{Which method finds explanations for more pairs?}
One notable difference in the evaluation measures is the explanation found rate, which assesses the number of user-item pairs for which an explanation can be found. As can be seen in Table \ref{tab:amzres2}, in general, the countERText technique performed the best, with an Explanation Found Rate of 98.9\% for cell-phones and 98\% for CDs. The Genetic and Gumbel techniques also performed well, but with lower Explanation Found Rates. This discrepancy is understandable, as the Gumbel-softmax algorithm requires changing words to other, semantically appropriate words, which can make it more difficult for the explanation generation method to find explanations. In contrast, the genetic and countERText methods only need to remove some words from the text, which is a simpler task. However, it is worth noting that the Gumbel-softmax method is designed to work well with datasets that contain a variety of feature types, such as textual, categorical, and numerical features. Therefore, it should not be expected to perform particularly well on a dataset that only contains textual features. Despite this, the Gumbel-softmax method still outperforms the baseline in both datasets.
\subsubsection{Stability}
It is shown in Table \ref{tab:amzres2} that, the baseline and countERText methods are the most stable on both datasets. The Genetic method has low stability on both datasets, with a stability score of 0.175 on the cell-phones dataset and 0.47 on the CDs dataset. The Gumbel method also has low stability, with a stability score of 0.51 on the cell-phones dataset and 0.54 on the CDs dataset.

The low stability scores for the Genetic and Gumbel methods could be due to the fact that these methods use randomness in their algorithms. For example, the Genetic method uses a genetic algorithm that involves random mutations and crossovers, which can lead to different results across runs. Similarly, the Gumbel method involves generating random variables and using them to calculate a probability distribution, which can also result in different results across runs. Moreover, the variance in the results across different runs could be attributed to the types of features utilized by each method. The baseline method employs a limited set of features, which results in less variance across runs. Conversely, the other methods utilize words as features, which can result in a greater degree of variability.  However, in the context of recommender systems, it may be beneficial to obtain different sets of features in different runs to provide users with diverse explanations from multiple perspectives. Thus, in this scenario, it would not be appropriate to determine whether a higher or lower stability is better.

\subsubsection{Time Complexity}
We evaluated the proposed method by measuring the average time spent generating explanations for a user-item pair. The time taken is reported in seconds in Table \ref{tab:amzres2}. For the cellphone and CD datasets, which only have textual features, the time spent generating explanations using our proposed methods depends entirely on the length of the textual descriptions, titles, and features of each item. As can be seen from our three proposed approaches, Gumbel-softmax takes the most time, while countERText takes the least time. However, the baseline method takes almost the same amount of time as countERText and Genetic algorithm. This is because the number of words used as features in our proposed algorithms is greater than the number of aspects used in the baseline method, and the search space is larger than that of the baseline method. Therefore, this difference is logical.
\subsection{Yelp dataset Results}
For the Yelp dataset, we conduct three experiments to investigate the effect of various types of features on the accuracy of the black-box model. Additionally, we examine the combination of multiple features on the explanations generated by the model. The experiments aim to determine the most effective features and feature combinations for both accurate predictions and clear explanations of the model's decisions.\\In the first experiment, we use all features, including textual features (extracted from users' tips), categorical features, and continuous features, to train a black-box model. To find explanations, we combine Equations \ref{eq3} and \ref{eq7} to define a new loss function, which can be used to find explanations for all types of features mentioned. Specifically, Equation 3 is used for continuous features and Equation \ref{eq7} is used for textual and categorical features. The second experiment focused solely on textual features and compared various methods for finding explanations, as previously conducted in the experiments on the Amazon dataset in Section \ref{amzresult}).The final experiment employed only categorical and continuous features to train the black-box model, and explanations were generated using the same methodology as in the first experiment.
The results of our experiments for the Yelp Dataset can be seen in Tables \ref{tab:yelpres} and \ref{tab:yelpres2}.
\begin{table}[]
	\centering
	\caption{Yelp Dataset Results, user-based and model-based}
	\label{tab:yelpres}
	\resizebox{\textwidth}{!}{%
			\begin{tabular}{c c ccc ccc}
			\hline
			Feature Types & &\multicolumn{3}{c}{User-based} & \multicolumn{3}{c}{Model-based} \\ 
			& & Pre & Rec & F1 & PN & PS & FNS \\ \hline
			\multirow{1}{*}{All} & Gumbel &\textbf{0.101} & 0.059 & 0.044 & 0.607 & 0.609 &  0.608 \\ \hline
			\multirow{4}{*}{Textual} & baseline & 0.080 & \textbf{0.1661} & \textbf{0.0909}  & 0.9544 &  \textbf{0.9619} &  0.9582  \\
			& countERText & 0.092 & 0.058 & 0.051 & 0.8966 & 0.9385 & 0.9171 \\
			& Genetic &  0.099 &  0.101 & 0.072 & \textbf{0.9712} &  0.956 & \textbf{0.9638} \\ 
			& Gumbel &  0.072 & 0.040 & 0.041 & 0.75 & 0.918 & 0.826  \\ \hline
			\multirow{1}{*}{Non-Textual} & Gumbel &  N/A & N/A & N/A & 0.647 & 0.495 & 0.555  \\ \hline
		\end{tabular}
	}
\end{table}
\begin{table}[]
	\centering
	\caption{Yelp Dataset Results, Other measures}
	\label{tab:yelpres2}
	\resizebox{\textwidth}{!}{%
		\begin{tabular}{c c c c c c c}
			\hline
		    Feature Types &  & NDCG & Features Avg & Exp Found Rate& Stability& Time(sec) \\ \hline
			\multirow{1}{*}{All} & Gumbel & \textbf{0.8499} & \textbf{7.6} & \textbf{98.3\%} & 0.59 & 5.4 \\ \hline
			\multirow{4}{*}{Textual} & baseline & 0.515 & \textbf{7.51} & 68\% & \textbf{0.74}& 1.36\\
			& countERText & \textbf{0.6172} & 13.819 & \textbf{93.4\%} & 0.58  & 0.78 \\ 
			& Genetic & \textbf{0.6172} & 17.24  & 73.76\% & 0.21 & 4 \\ 
			& Gumbel & \textbf{0.6172} & 11.39 & 59.2\% & 0.51 & 2.8\\ \hline
			\multirow{1}{*}{Non-Textual} & Gumbel & \textbf{0.8434} & \textbf{2.73} & 89.9\% & \textbf{0.77} & 2\\ \hline
		\end{tabular}
	}
\end{table}
\subsubsection{Are generated features based on user preferences?}
As can be observed in Table \ref{tab:yelpres}, the results of evaluating features based on user preferences are consistent with those obtained from the Amazon datasets. Given that user-preference ground-truth features are extracted from textual features, it is not possible to evaluate methods that do not utilize textual features. By comparing the results for textual features, it can be inferred that the baseline method achieves better results in this aspect, as fewer features need to be identified by the method. The analysis of results in this section aligns with the findings presented in Section \ref{amzresult}.
\subsubsection{Are generated features specifying the model's behavior?}
As can be seen in Table \ref{tab:yelpres}, when only textual features are used, the generated explanations provide a better understanding of the model's behavior. However, when using non-textual features, the generated explanations are not as effective in specifying the model's behavior. It may be due to the fact that many of the non-textual features may be unrelated to certain businesses, and thus their removal from the list of features does not significantly impact the ranking score of these businesses. As a result, the PN score is lower than that of other methods. Similarly, for the PS score, the same phenomenon occurs, where some features are unrelated to certain businesses and are thus insufficient to place them in the top K list, resulting in a lower score compared to other methods. One potential solution to this issue is to consider categorical features related to each type of business separately.

Additionally, when all features are utilized, the results are better than using only non-textual features but not as favorable as when only textual features are used. This is likely because non-textual features are included among the textual features. 

Furthermore, when utilizing textual features, the genetic algorithm performed the best, but with a larger number of features found. Moreover, the baseline and countERText methods were the next best performers, while the Gumbel method performed the worst.

In conclusion, our contribution is the ability to find explanations that contain multiple types of features simultaneously through the Gumbel-softmax method. By improving the formulation of categorical and continuous features, it is hoped that the results will further improve.
\subsubsection{The effect of changing features on ranking}
As can be seen in Table \ref{tab:yelpres2}, when all features are employed or only categorical and continuous features are utilized, the NDCG value is more favorable. Conversely, the utilization of textual features resulted in a lower NDCG value for the black-box model. This can be attributed to the fact that the textual features used in this study were extracted from users' tips on the items, rather than from the items' features directly. This trend is consistent with the results observed when using user reviews to extract textual features for the Amazon dataset. In that case, we use the item's descriptions and features directly instead of user reviews, to avoid this issue. Nonetheless, non-textual features demonstrated a noteworthy NDCG score. By comparing the methods used for extracting textual features, it can be inferred that the method of feature extraction plays a crucial role in the NDCG score of the black-box model.
\subsubsection{Which method finds explanations for more pairs with lower number of features in average?}
As can be seen in Table \ref{tab:yelpres2}, when all features are used, the explanation generation algorithm is able to generate explanations for 98\% of pairs with an average of 7.6 features. This high success rate is achieved despite the need to search through a large number of features, indicating that the algorithm is able to generate explanations for a significant proportion of pairs with relatively few features. On the other hand, the baseline method which only utilizes textual features generates explanations for only 68\% of pairs with an average of 7.51 features. The countERText method performs well with 93.4\% of explanation found rate but with a large number of features in average (13.81). The genetic algorithm also performs well but with more features in average 17.24. The Gumbel softmax method, however, performs relatively poorly, generating explanations for only 59.2\% of pairs with an average of 11.39 features. Utilizing only non-textual features results in generating explanations for 89.9\% of pairs with an average of 2.73 features.
Overall, it appears that the Gumbel method when using all features is able to generate explanations for a high proportion of pairs with a relatively low number of features on average.
\subsubsection{Stability}
As can be seen in Table \ref{tab:yelpres2}, it becomes apparent that the stability of the baseline method is 0.74 when only textual features are used, which is higher than that of the other methods, consistent with the findings from the Amazon dataset.
On the other hand, the Gumbel method using all types of features shows a stability of 0.59, which is higher than when only textual features are used. This suggests that the features generated by the Gumbel method are more consistent when all types of features are used. When non-textual features are used, the stability is higher than the other methods. This can be attributed to the fact that the number of non-textual features is much smaller than the number of textual features. 
These results highlight the importance of selecting appropriate features for a given task and using a robust method to generate stable features.
\subsubsection{Time Complexity}
The time spent generating explanations for the Yelp dataset is shown in Table \ref{tab:yelpres2}. This time not only depends on the length of tips used as textual features but also on the number of categorical features we have. However, since the tips in the Yelp dataset are generally shorter than the descriptions, features, and titles in the Amazon dataset, the time spent generating explanations for only textual features is less than that for the Amazon dataset on average. However, as other categorical features are added and all types of features are used, the time spent generating explanations increases. Nevertheless, as with the Amazon dataset, the feature space in which we search for generating counterfactual explanations for our methods is larger than that of the baseline method, so the difference in the time spent generating explanations is logical.
\begin{table}[]
	\caption{cell phones dataset example 1}
	\label{tab:cellsEx}
	\begin{tabular}{|l|l|}
		\hline
		user\_id &
		A3VVMIMMTYQV5 \\ \hline
		item\_id &
		B00U7YKO78 \\ \hline
		baseline &
		phone, battery, lights \\ \hline
		countERText &
		\begin{tabular}[c]{@{}l@{}}removing words: 'version','international'\\ ,'warranty','samsung','wireless','us'\end{tabular} \\ \hline
		Genetic &
		\begin{tabular}[c]{@{}l@{}}removing words: 'qi','galaxy','note', 'go','select',\\ 'daydream','more','corner','samsung','international',\\ 'us',devices',' micro','usb','charging','wpc'\end{tabular} \\ \hline
		Gumbel &
		\begin{tabular}[c]{@{}l@{}}words : {[}'cover', 'samsung', 'wireless', 'international', 'white'{]},\\ change to: {[}'daydream', 'with', 'galaxy', 'board', ';'{]}\end{tabular} \\ \hline
		\begin{tabular}[c]{@{}l@{}}user review\\ on item\end{tabular} &
		\begin{tabular}[c]{@{}l@{}}{[}'plastic', 'back', 'I have a Spigen case that has a clear plastic back'{]},\\  {[}'charge','quick',"It charged so quick I didn't get the chance to really\\  figure out when it topped off"{]},\\  {[}'phone', 'right', 'I charged my phone right away to try it out'{]},\\  {[}'case', 'clear',  'I have a Spigen case that has a clear plastic back'{]},\\  {[}'case', 'clear', 'My Spigen case was a clear plastic'{]},\\  {[}'lights', 'blue',  'the blue lights will start blinking'{]},\\  {[}'lights','blue','it emits 2 deep blue tubular lights along the front side\\  to let you know\end{tabular} \\ \hline
	\end{tabular}
\end{table}
\begin{table}[]
	\caption{cell phones dataset example 2}
	\label{tab:cellsEx2}
	\begin{tabular}{|l|l|}
		\hline
		user\_id &
		A1F7YU6O5RU432 \\ \hline
		item\_id &
		B00Z7RQ0NC \\ \hline
		baseline &
		\begin{tabular}[c]{@{}l@{}}phone, quality, buttons,device, case, protection, cases,screen\\ ,color, grip\end{tabular} \\ \hline
		countERText &
		\begin{tabular}[c]{@{}l@{}}removing words:'lifetime','date','protects','graphic','plus','iphone'\\ ,'absorbs','seamless','night','withstands','edge','white'\end{tabular} \\ \hline
		Genetic &
		\begin{tabular}[c]{@{}l@{}}removing words: 'phone','iphone','amp','iphone','only','date'\\ ,'iphone','details'\end{tabular} \\ \hline
		Gumbel &
		no explanation found \\ \hline
		\begin{tabular}[c]{@{}l@{}}user review\\ on item\end{tabular} &
		\begin{tabular}[c]{@{}l@{}}{[}'case','pretty','Love the looks of it and when\\  the sun catches the silver flecks on the case its so pretty'{]},\\  {[}'case','clear','The case is clear so whatever color your iPhone\\  is youll be able to see it a bit through the sparkles'{]},\\  {[}'photos','online', 'What might be hard to tell initially from the\\  online photos is this case has a bunch of gorgeous sparkles inside\\  the clear plastic\textbackslash{}'\end{tabular} \\ \hline
	\end{tabular}
\end{table}
\begin{table}[]
	\caption{CDs dataset example 1}
	\label{tab:cdsEx}
	\begin{tabular}{|l|l|}
		\hline
		user\_id &
		A3LEN0P07MGJE2 \\ \hline
		item\_id &
		B001TRDPB4 \\ \hline
		baseline &
		no explanation found \\ \hline
		countERText &
		removing words: 'christmas', 'cheers \\ \hline
		Genetic &
		\begin{tabular}[c]{@{}l@{}}removing words: 'members','living','sell','sensation','2008'\\ ,'nationwide','tour','bowl','bring','another','chaser','jingle',\\ 'christmas','cheers'\end{tabular} \\ \hline
		Gumbel &
		\begin{tabular}[c]{@{}l@{}}words : {[}'cheers', 'christmas', 'reindeer'{]}\\ change to: {[}"'", 'filmed', 'tonight', '!'{]}\end{tabular} \\ \hline
		\begin{tabular}[c]{@{}l@{}}user review\\ on item\end{tabular} &
		\begin{tabular}[c]{@{}l@{}}{[}{[}'twists','new','There were some songs that were new to me \\ as well as new twists to old favorites -- the introduction to \\ We Three Kings comes to mind with its hints of the\\  Mission Impossible theme'{]},\\ {[}'twists', 'in','The twists in Rudolph the Red-Nosed Reindeer'{]}{]}\end{tabular} \\ \hline
	\end{tabular}
\end{table}
\begin{table}[]
	\caption{CDs dataset example 2}
	\label{tab:cdsEx2}
	\begin{tabular}{|l|l|}
		\hline
		user\_id    & A1SCJWCMQ3W3KK                                                                                                                \\ \hline
		item\_id    & B00006879E                                                                                                                    \\ \hline
		baseline    & song , rock, blues, tune, release, disc, collection, hit                                                                      \\ \hline
		countERText & removing words: 'jane'                                                                                                        \\ \hline
		Genetic     & removing words: 'first', 'songs'                                                                                              \\ \hline
		Gumbel      & \begin{tabular}[c]{@{}l@{}}words:{[}'stevie', 'jane', 'songs'{]}, \\ change to:{[}'their', 'rhythms', 'those'{]}\end{tabular} \\ \hline
		\begin{tabular}[c]{@{}l@{}}user review\\ on item\end{tabular} &
		\begin{tabular}[c]{@{}l@{}}{[}'release', 'in',"while 'Tangled' is more reminiscent of something\\  Motown would have released back in the 1970s"{]}, \\ {[}'release', 'back', 1, "while 'Tangled' is more reminiscent of \\ something Motown would have released back in the 1970s"{]},\\  {[}'music', 'fresh', 'the record still holds up as fresh music'{]}, \\ {[}'band', 'in', 'which takes the band into a slightly softer feel'{]}\\ {[}'blues', 'little', 'a little blues'{]}\\ {[}'rock', 'in', 'and rock and roll in a way that was previously\\  unknown (at least on a grand scale) on US airwaves'{]}\\ {[}'rock', 'little', 'a little rock'{]},\\ {[}'tracks', 'first',  'the first eight tracks are spectacular'{]}\\ {[}'funk', 'little', 'A little funk\textbackslash{}'{]}\end{tabular} \\ \hline
	\end{tabular}
\end{table}
\subsection{Qualitative Evaluation} \label{qualitativeEv}
In this section we present some examples of features found by different methods for different datasets. \\The examples of the cell-phones dataset are presented in Tables \ref{tab:cellsEx} and \ref{tab:cellsEx2}. Due to the length of the user reviews, only the ground-truth aspects and the sentences containing them have been included. It should be noted that without access to the full reviews, it may not be possible to fully evaluate the effectiveness of each method. Nevertheless, as observed in the first example, the word 'charge' is an aspect identified by the genetic algorithm and mentioned in the user review. Additionally, the words 'case' and 'plastic' are also found in the user review and are similar to the aspect 'cover' identified by the Gumbel method. The aspect 'light' is also identified by the baseline method as being important in the user reviews. However, for a fair comparison, it is necessary to have access to the full descriptions and features of the item and the entirety of the user reviews to understand the context in which these words are used and the specific meaning they convey. Due to the length of these texts, they have not been included in this presentation.

The examples of the CDs dataset are presented in Tables \ref{tab:cdsEx} and \ref{tab:cdsEx2}. As can be seen, the baseline method did not identify any explanations in the first example, while the other methods found some explanations. However, as with the cell-phones dataset, it is not possible to determine the relevance of these identified aspects without access to the full reviews and descriptions. In the second example, it is observed that the user expressed a preference for the first tracks and the words "first" and "songs" are identified as important by the genetic algorithm. Moreover, the user mentions the words "rock", "blue", and "release" in their review and these words are identified by the baseline method. The Gumbel method and CountER both identified a name as an important aspect, which may be relevant to the user but may also be found in other parts of the review that are not included in this table.
\begin{table}[]
	\caption{Yelp dataset example 1}
	\label{tab:my-Yelpex}
	\begin{tabular}{|cl|l|}
		\hline
		\multicolumn{2}{|c|}{\multirow{2}{*}{user\_id}} &
		\multirow{2}{*}{nlReKgQoRz6uPfVaEG93mw} \\
		\multicolumn{2}{|c|}{} &
		\\ \hline
		\multicolumn{2}{|c|}{item\_id} &
		tU692E8N0xBQ7Ogc78gN2g \\ \hline
		\multicolumn{2}{|c|}{all features} &
		\begin{tabular}[c]{@{}l@{}}
			latitude  change from 36.177038 to  36.17646826\\  longitude  change from  -86.749691 to -86.75116574 \\ words :  {[} 'going', 'anything', 'rain', 'potato', 'house'{]} \\
			change to : {[} 'about', 'spot', 'much', '!', 'black'{]}	
		\end{tabular} \\ \hline
		\multicolumn{1}{|c|}{\multirow{4}{*}{\begin{tabular}[c]{@{}c@{}}textual \\ features\end{tabular}}} &
		baseline &
		\begin{tabular}[c]{@{}l@{}}	staff, taste, inside \end{tabular} \\ \cline{2-3} 
		\multicolumn{1}{|c|}{} &
		\begin{tabular}[c]{@{}l@{}}countER\\ Text\end{tabular} &
		\begin{tabular}[c]{@{}l@{}}removing words:{[}'milky','going','way','table','coffee',\\ 'place','great','mocha','try','yummy','numb','best'\\ ,'nashville','breakfast','amazing','hangout','get'{]}\end{tabular} \\ \cline{2-3} 
		\multicolumn{1}{|c|}{} &
		Genetic &
		\begin{tabular}[c]{@{}l@{}}removing words:{[}'order','feastival','addictive','green',\\ 'love','atmosphere','milky','best','town','joe','everything'\\ ,'wrong','brazilian','fuzzy','bomb','fishy','ever','roasted',\\ 'great','breakfast','simple','syrup','iris','flower','cream',\\ 'sliced','table','hangout','bloom','iced','americano',\\ 'grapefruit','tea','cucumber','sweet','super','latte',\\ 'office','bar','pizza','apple','sauce', 'decent','coffee'{]}\end{tabular} \\ \cline{2-3} 
		\multicolumn{1}{|c|}{} &
		Gumbel &
		\begin{tabular}[c]{@{}l@{}}no explanation found \end{tabular} \\ \hline
		\multicolumn{2}{|l|}{\begin{tabular}[c]{@{}l@{}}non-textual features\end{tabular}} &
		\begin{tabular}[c]{@{}l@{}}latitude change from 36.177038 to 36.17698086\\ longitude change from -86.749691 to -86.7495435 \end{tabular} \\ \hline
		\multicolumn{2}{|l|}{\begin{tabular}[c]{@{}l@{}}user review on business\end{tabular}} &
		\begin{tabular}[c]{@{}l@{}}It's an awesome place to drop in and eat or get something to go\end{tabular} \\ \hline
	\end{tabular}
\end{table}

\begin{table}[]
	\caption{Yelp dataset example 2}
	\label{tab:my-Yelpex2}
	\begin{tabular}{|cl|l|}
		\hline
		\multicolumn{2}{|c|}{\multirow{2}{*}{user\_id}} &
		\multirow{2}{*}{0du93EkEwKuxRG\_x6hqVUg} \\
		\multicolumn{2}{|c|}{} &
		\\ \hline
		\multicolumn{2}{|c|}{item\_id} &
		KnsY8rh5tigp5t6WpilGdA \\ \hline
		\multicolumn{2}{|c|}{all features} &
		\begin{tabular}[c]{@{}l@{}}latitude: change from 36.103133 to 36.10256344\\ longitude: change from -86.8185 to -86.81997484  \\ Open24Hours: change from Not-mentioned to True\\ words: {[}'bars','gallon','sunday','bag','girl', 'sausage',\\ 'breakfast','better','awesome'{]}\\ change to: {[}'chef', '?', 'red', 'guy', 'salad', 'foods', \\ 'you', 'start', 'dawn'{]}\end{tabular} \\ \hline
		\multicolumn{1}{|c|}{\multirow{4}{*}{\begin{tabular}[c]{@{}c@{}}textual \\ features\end{tabular}}} &
		baseline &
		\begin{tabular}[c]{@{}l@{}}coffee, favorite, spot, cheese, tasting,\\ price, eating, hour, ingredients, tea,\\ chocolate, neighborhood, shop\end{tabular} \\ \cline{2-3} 
		\multicolumn{1}{|c|}{} &
		\begin{tabular}[c]{@{}l@{}}countER\\ Text\end{tabular} &
		\begin{tabular}[c]{@{}l@{}}removing words:{[}'mean',class','includes','day','bar','run',\\ 'nashville','buffet','hour','hills','butter',\\ 'chai','guys','rush','come','food','beer',\\ 'thanksgiving','see','rules','pick','today','salad{]}\end{tabular} \\ \cline{2-3} 
		\multicolumn{1}{|c|}{} &
		Genetic &
		\begin{tabular}[c]{@{}l@{}}removing words:{[}'bags','juice', 'lunch','run','variety', 'juice',\\ 'traditional','thanksgiving','feel','shopping','girl','thank','nice','butter',\\ 'meat','watch','hockey','pork','wonderful','act','ordered',\\ 'took','come','salad','get','thai','pump','saturdays','think'\\ ,'come','hidden','bar','almond','including','beer','narragansett',\\ 'hot','figured','delicious','desert','breakfast','consistent','option',\\ 'selection','selection','breakfast','chicken','hot','rules',\\ 'looking','epic','grilled','nice','pepper','guys','great','organic',\\ 'pizza','order','table','awesome','orange','makings','free{]}\end{tabular} \\ \cline{2-3} 
		\multicolumn{1}{|c|}{} &
		Gumbel &
		\begin{tabular}[c]{@{}l@{}}words:{[}'southern','saturdays','best','hills','locations','rules',\\ 'bar','guys', 'today','wonderful','awesome'{]}\\ change to:\\ {[}'noon','was','any','village',';', 'did','also','sausage',\\  'like','get','.','to','eat','little'{]}\end{tabular} \\ \hline
		\multicolumn{2}{|l|}{\begin{tabular}[c]{@{}l@{}}non-textual\\  features\end{tabular}} &
		\begin{tabular}[c]{@{}l@{}}latitude: change from 36.103133 to 36.10318994\\ longitude: change from -86.8185 to -86.81864752  \\ halal: change from to Not-mentioned to False\end{tabular} \\ \hline
		\multicolumn{2}{|l|}{\begin{tabular}[c]{@{}l@{}}user review\\ on business\end{tabular}} &
		\begin{tabular}[c]{@{}l@{}}{[}{[}'cheese', "You'll find interesting cheese"{]},\\  {[}'chicken salad', "and awesome chicken salad"{]},\\  {[}'foods',  'Their prepared foods are also pretty awesome'{]}{]}\end{tabular} \\ \hline
	\end{tabular}
\end{table}
Examples for the Yelp dataset are presented in Tables \ref{tab:my-Yelpex} and \ref{tab:my-Yelpex2}. As demonstrated, when utilizing all types of features, the Gumbel method can identify changes in continuous, categorical, and textual features. For instance, in the first example, there are slight modifications in the wording of the tips and the latitude and longitude of the business, the establishment is no longer recommended to the user. \\
Furthermore, the second example illustrates that when utilizing only non-textual features, it is discovered that if the halal option is changed from "not mentioned" to "False," indicating that the restaurant does not serve halal food, the establishment is no longer recommended to the user.\\
An analysis of user reviews for the business reveals that the words 'chicken', 'salad' and 'food' are identified as important by various explanation methods, indicating their results are acceptable. 

It is noteworthy that for each business, changes in categorical features can be identified separately, and recommendations can be made to the establishment to consider these features. For example, informing the manager that serving halal food or providing bicycle parking may attract more customers.
\section{Conclusion and Future works}\label{conclusion}
In this paper, we introduced three methods for generating explanations for textual explanations and evaluated them on three real-world datasets of recommender system tasks. CountERText and Genetic methods were able to find only textual features as explanations, while the Gumbel method, which employed Gumbel softmax, was able to be applied to all types of features, including textual, categorical, and continuous features. We conducted experiments to evaluate these methods and found that when item textual features were used, our method outperforms the baseline in terms of model-based measures, meaning that the features found as explanations were both necessary and sufficient for the model to make accurate predictions. Although the models did not perform well when using user tips on items as textual features, the Gumbel softmax-based method has the potential to produce explanations based on multiple features, which could be useful for tasks involving multi-modal features such as healthcare.

In future works, the model can be improved by generating explanations for different types of businesses separately, so that their categorical features are not combined. Additionally, one could consider only nouns as textual features and generate explanations based on them. Furthermore, the way of evaluating explanations based on user preferences can be improved by considering the semantic similarity of user reviews and all found features.
%
%
\bibliographystyle{spbasic}      

\bibliography{references}


%
%

\end{document}